\documentclass[english,a4paper,aps,prl,twocolumn]{revtex4}
\usepackage{amsmath}
\usepackage{amssymb}
\usepackage{graphicx}

 \usepackage{amsbsy}
\usepackage{babel}
\begin{document}
\title{Imaging of double slit interference by scanning gate microscopy}
\author{K. Kolasi\'nski, B. Szafran, and M.P. Nowak}
\address{AGH University of Science and Technology, \\ Faculty of Physics and Applied Computer Science, \\ al. A. Mickiewicza 30, 30-059 Krakow, Poland}
\begin{abstract}
We consider scanning gate microscopy imaging of the double slit interference for a pair of quantum point contacts (QPCs) defined within the two-dimensional electron gas.
The interference is clearly present in the scattered electron wave functions for each of the incident subbands.
Nevertheless, we find that the interference is generally missing in the experimentally accessible conductance maps
for many incident subbands. We explain this finding on the basis of the Landauer approach.
 A setup geometry allowing for observation of the double slit interference by the scanning gate microscopy is proposed.
\end{abstract}

\maketitle

\section{ Introduction} The interference effects for particles \cite{dg} are the cornerstone
of quantum mechanical understanding of the wave nature of matter.
The experiments including  double slit interference  were
performed with electrons \cite{dse,hass}, ions \cite{cronin} and larger objects including clusters and molecules \cite{horn}.
The double-slit interference implementing the  Feynman version of the experiment was reported only very recently \cite{bach}.
The interference experiments are performed on particles in vacuum \cite{dg,dse,hass,cronin,horn} or in the solid state. 
The two-dimensional electron gas (2DEG) in semiconductor nanostructures is an attractive solid-state medium in this context due to a large coherence
length for Fermi level electrons and carrier confinement that can be arbitrarily tailored for formation of electron interferometers \cite{ei}.
The role of slits in 2DEG is played by quantum point contacts (QPCs)  \cite{yacoby,khatua}.

The electron flow determined by the Fermi level wave function is imaged
by the scanning gate microscopy (SGM) \cite{sgmr}. In the SGM technique the charged tip of the atomic force
microscope locally perturbs the potential landscape within the 2DEG which is buried shallow beneath the surface of the semiconductor.
 The response of the system is monitored by conductance ($G$) maps as functions of the tip position.
SGM was early used in quantum point contacts \cite{qpc}
for observation of the wave function diffraction including angular branching due to the conductance quantization
inside the slit. The conductance maps contain interference pattern involving electron standing waves formed between the QPC
and the tip \cite{jura}. SGM imaging of electron interferometers using antidots was also performed \cite{hackens}.
In spite of a large number of SGM studies of a single QPC \cite{qpc,jura,th}
the imaging of the double slit interference was not reported so far. This paper investigates
the problem of observation of the double slit interference imaging by SGM.

We solve the phase coherent scattering problem for the Fermi level electron waves incident to the double slit system.
The calculated SGM conductance map for the simplest double slit system does {\it not} contain signatures of the Young interference
and is a sum of $G$ maps for separate slits.
The double slit interference is present in the contributions to conductance for separate incident subbands
but it disappears in the Landauer \cite{datta} summation. We explain this finding using symmetry arguments
and indicate the setup geometry needed to obtain the interference signal in the SGM conductance map.

\begin{figure*}[htbp]
\begin{centering}
\includegraphics[width=0.85\paperwidth]{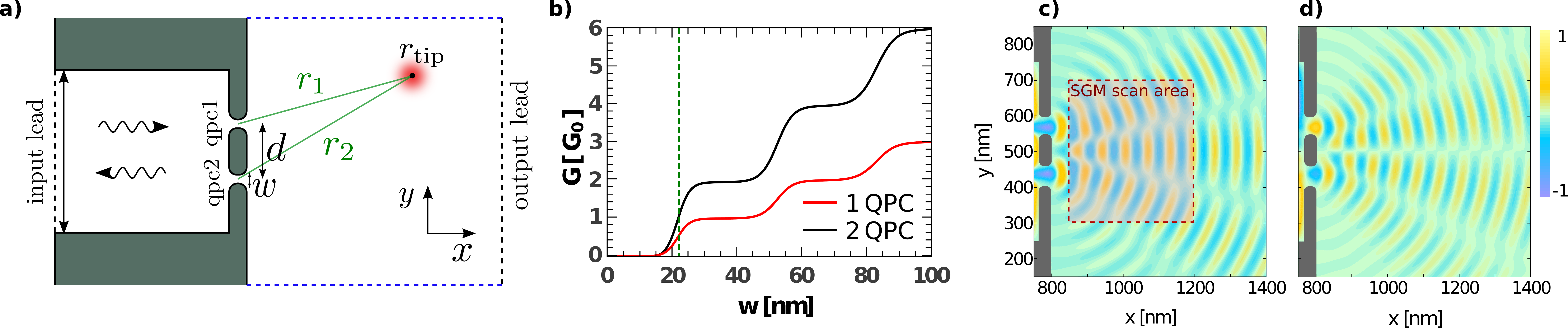}
\par\end{centering}
\caption{\label{sch}a) Sketch of the double-slit system for the electron incident
from the left (lead of width 0.5$\mu$m) with two QPCs
of width $w$ nm with centers separated by $d=130$ nm. The space at right of the slits is infinite with transparent boundary
conditions \cite{nowak1} applied along the dashed lines \cite{actual}. $r_i$ is the distance
between $i$-th QPC and the tip. b) Conductance for a single (red line) and both QPCs (black line) open
for Fermi energy $E_F=6$ meV (16 conducting subbands in the input lead). c) Real part of the wave function for the electron incident
from the lowest (c)
and second (d) subband.  }
\end{figure*}

\section{Theory}
We consider a system depicted in Fig. \ref{sch}(a) with a wide (500 $\mu$m) input lead and two open slits. 
We solve the coherent scattering problem  as given
by the Schr\"odinger equation for the Fermi level electrons.  In
order to determine the conductance we use a finite difference variant \cite{kolas}
of the quantum transmitting boundary method \cite{Lent90,Lent94}
for the effective mass Hamiltonian \begin{equation} H=-\frac{\hbar^{2}}{2m}{\nabla}^{2}+V_\mathrm{con}(x,y)+V_{\mathrm{tip}}(x,y)\end{equation}
where $m=0.067 m_0$ is the GaAs electron effective mass,  $V_\mathrm{con}$ and $V_\mathrm{tip}$ are the potentials forming
the confinement (a finite quantum well is applied) and describing the perturbation induced by the scanning tip, respectively.
For the tip potential we use the Lorentz function as obtained in our previous Schr\"odinger-Poisson modeling \cite{szafran2011},
\begin{equation} V_{\mathrm{tip}}(x,y)=\frac{U_{\mathrm{tip}}}{1+\left[\left(x-x_{\mathrm{tip}}\right)^{2}+\left(y-y_{\mathrm{tip}}\right)^{2}\right]/d_{\mathrm{tip}}^{2}},\end{equation}
where $(x_\mathrm{tip},t_\mathrm{tip})$ is the position of the tip, $d_\mathrm{tip}$ is the width of the potential maximum
and $U_\mathrm{tip}$ its value.

In the input lead far from the QPCs the wave function is a superposition of incoming and backscattered waves
\begin{equation} \Psi^{\mathrm{input}}(x,y)=\sum_{k=1}^{M}a_{k}e^{ikx}\chi_{k}^{\mathrm{in}}(y)+b_{k}e^{-ikx}\chi_{-k}^{\mathrm{in}}(y)\end{equation}
where the summation runs over wave vectors $k$ at the Fermi level.
In the upper and lower ends of the computational box (the blue dashed lines in Fig. \ref{sch}) transparent boundary conditions are applied \cite{nowak1}.
In the output lead (black dashed line in Fig. \ref{sch}(a)) of width 1$\mu$m we have  the outgoing wave functions only
\begin{equation} \Psi^{\mathrm{output}}(x,y)=\sum_{k=1}^{M_{\mathrm{out}}}d_{k}e^{ikx}\chi_{k}^{\mathrm{out}}(y).\end{equation}
In this paper we take $E_F=6$ meV as the Fermi energy level for which we have $M=16$, and $M_\mathrm{out}=32$ conducting subbands
in the input and output leads, respectively. After solution of the scattering problem
for each incoming mode the conductance of the system is evaluated
by the Landauer formula $G=G_{0}\sum_{i=1}^{M_{\mathrm{in}}}T_{i}$,
where $T_{i}$ is the transmission probability of the $i$-th incoming mode and $G_0={2e^2}/{h}$

\section{Results and Discussion}
The calculated conductance for a single and two QPCs in the absence of the tip is plotted in Fig. \ref{sch}(b)
as a sum of contributions from 16 conducting subbands of the input lead. For further discussion we choose $w=22$ nm at the first step of $G$:
we shall discuss
the simplest case of a single conducting subband for each QPC. The conductance maps obtained by the SGM for the QPC tuned to the $G$ steps exhibits oscillations with a high amplitude and radial fringes \cite{jura}, which is convenient for discussion of the interference involving both slits.
The real part of the wave functions is displayed in Fig. \ref{sch}(c,d) for the transport calculated from the lowest and the second subband of opposite symmetry
with respect to the lead axis.

\begin{figure}[htbp]
\begin{centering}
\includegraphics[width=0.4\paperwidth]{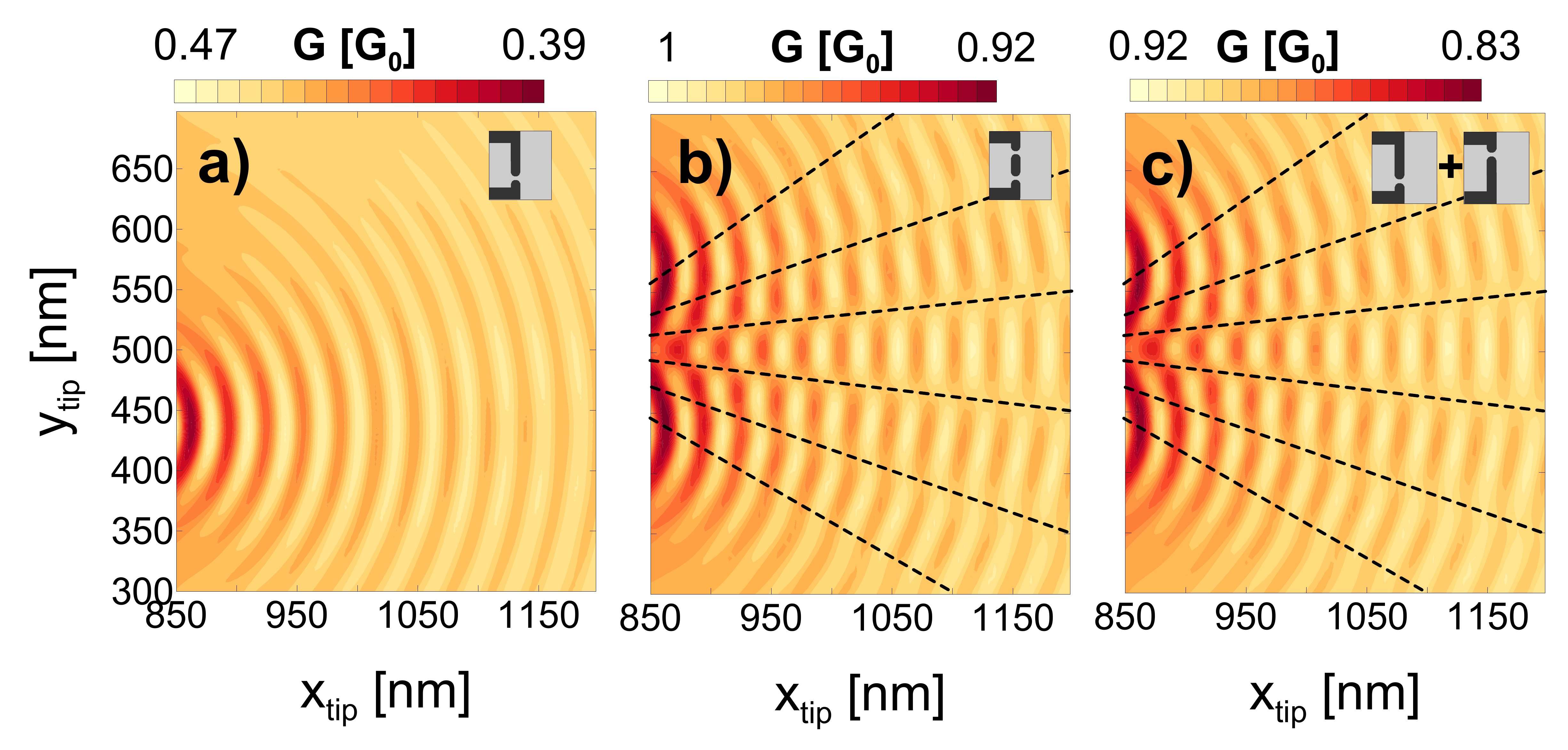}
\par\end{centering}
\caption{Conductance map for the QPC no. 2 open (a) and for both QPCs open (b). (c) Sum of conductance
maps for separate QPCs open for tip potential parameters: $d_{\mathrm{tip}}=10$ nm and $U_{\mathrm{tip}}=5$ meV. The insets schematically
explain the setup of the apparatus for each of the plots. \label{sgm}}
\end{figure}

The dashed rectangle of Fig. \ref{sch}(c) indicates the area in which we will discuss the conductance maps obtained by SGM.
The $G$ map calculated for a single QPC by the scattering problem with the method described in the previous section is displayed in Fig. \ref{sgm}(a).
The map contains the radial fringes due to formation of the standing waves
between the QPC and the tip. When both QPCs are open [Fig. \ref{sgm}(b)] we observe a checkerboard pattern which however
turns out {\it not} to be the effect of the double-slit wave function interference, as the pattern of the $G$ map can
be exactly reproduced by the sum of $G$ maps for separate open slits -- see Fig. \ref{sgm}(c).

\begin{figure}[htbp]
\begin{centering}
\includegraphics[width=0.4\paperwidth]{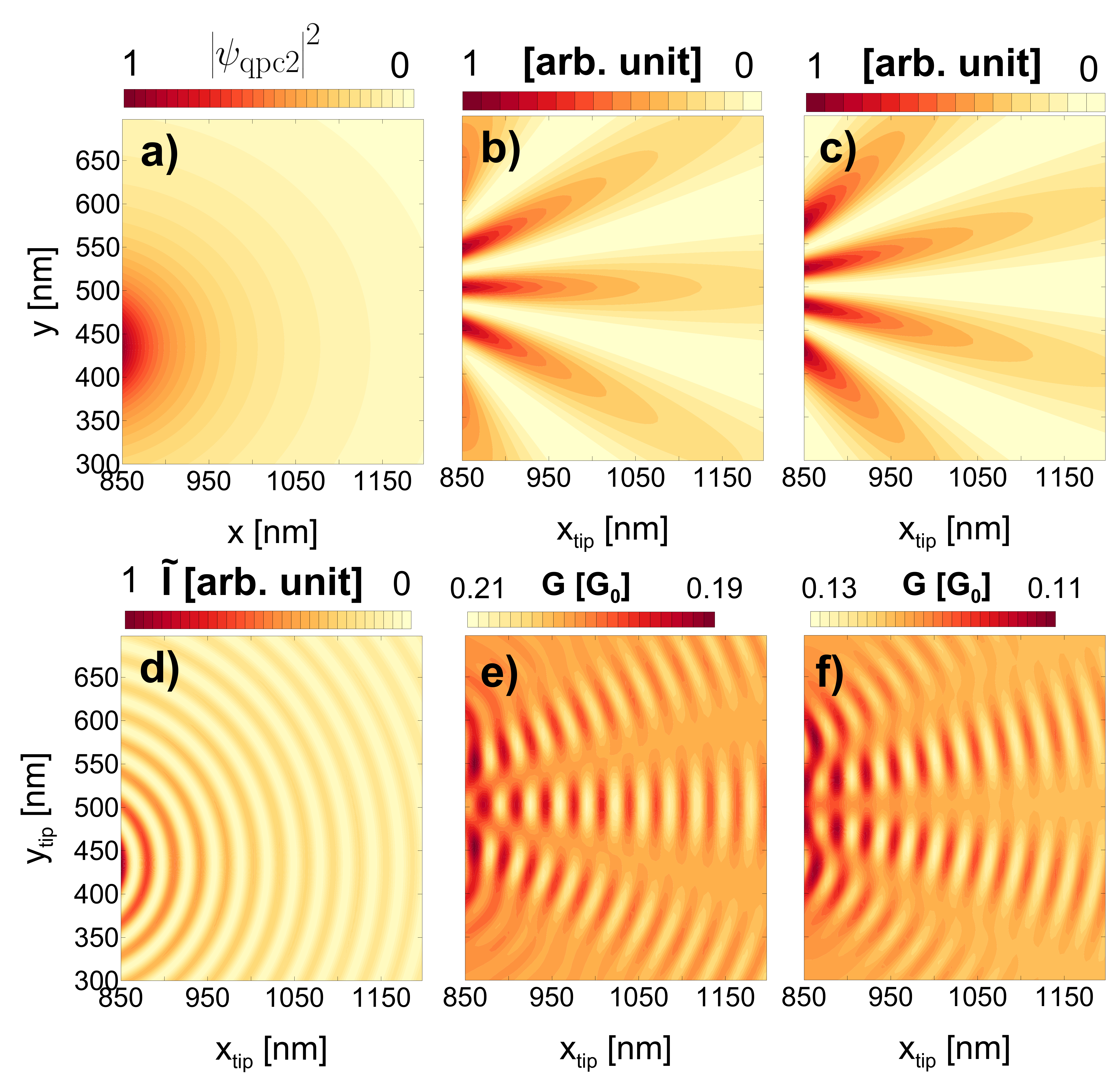}
\par\end{centering}
\caption{(a) Probability density for the diffracted wave function for the second QPC open in the absence of the tip
as calculated from the scattering problem. (b) and (c) show the probability densities for interference
of diffracted waves $|\Psi_1\pm \Psi_{2}|^2$ with '+' in (b) and '-' in (c).
(d) The variation of the probability density as given by Eq. (11) to be compared with the SGM map of Fig. 2(a). (e) and (f) show
contributions to conductance in function of the tip position for the lowest (e) and second (f) incident subband.  \label{dy}}
\end{figure}

Let us explain the absence of the Young interference in the conductance maps. We begin by the case of a single QPC open. In the absence of the tip the wave function passing
through the slit gets diffracted. 
The diffracted wave function can be quite well numerically simulated by the Huygens principle.
Each point of the opening is a source of a circular wave and the resulting diffracted wave function 
can be calculated as \begin{equation} \Psi_{{qpc}}(\boldsymbol{r})=\int_{\mathrm{qpc}} d {\bf r'}f_{\mathrm{qpc}}(y') \frac{e^{ik_F |r-r'|}}{\sqrt{|{\bf r-r'}|}} \end{equation}
with the Fermi wave vector at the open half-space side ($\frac{\hbar^2 k_F^2}{2m}=E_F$) and the transverse wave function of the only conducting mode, which
for the QPC center at the origin is $f_\mathrm{qpc}(y)=\sqrt{\frac{2}{w}}\sin(\frac{\pi y}{w})$ (for an infinite quantum well forming the constriction). The diffracted density
calculated by Eq.(5) is displayed in Fig. \ref{dy}(a).
The diffracted wave function can be put in form \begin{equation} \Psi_{l}(\boldsymbol{r})=\psi_{\mathrm{{qpc}}}({\bf r}) e^{i k_F r},\label{difr}\end{equation} where $\psi_{\mathrm{qpc}}$ is
a slowly varying envelope function.   When the tip is introduced to the system at position ${\bf r_{tip}}$
it gives rise to a scattering wave function $\Psi_{bs}$ with the amplitude proportional to the envelope $\psi_{\mathrm{qpc}}$,
\begin{equation} \Psi_{bs}({\bf r})=R({\bf r_{tip}-r}) \psi_{\mathrm{qpc}}({\bf r})e^{i{k_F}|{\bf {r_{tip}}-r}|}.\end{equation} The  $R$ function varies slowly as compared to the Fermi wavelength $2\pi/k_F$.
Along the line  between the tip and the QPC the superposition of the incoming [Eq. (6)] and backscattered [Eq. (7)] wave functions  is approximately given by
\begin{equation}\Phi({\bf r})=\psi_{\mathrm{qpc}}({\bf r})\Phi_{bs}({\bf r})\end{equation} with \begin{equation} \Phi_{bs}=(e^{ik_Fr_l}+\alpha e^{-ik_Fr_l}),\end{equation} where a scattering amplitude $\alpha$ replaces $R$ function and $r_l$ stands for the distance to the  $l$-th slit [see Fig. \ref{sch}(a)].
The SGM image is proportional to the probability density  \begin{eqnarray} I_{SGM}({\bf r})&=&|\Phi({\bf r})|^2=|\psi_{\mathrm{qpc}}({\bf r})\Phi_{bs}({\bf r})|^2= \\&&|\psi_{\mathrm{qpc}}({\bf r})|^2 \left(|\alpha^2|+1+2\Re{(\alpha)}(\cos^2 (k_Fr_l)-1)\right). \nonumber\end{eqnarray}
Neglecting the slowly varying terms the spatial variation of the probability density distribution is given by \begin{equation} \tilde{I}_{SGM}=|\psi_{\mathrm{qpc}}({\bf r})|^2\cos^2(k_Fr_l).\end{equation}
Figure \ref{dy}(d) shows $\tilde{I}$ for the 2nd QPC open which  very well coincides with the variation of the exact conductance map of Fig. 2(a) with the oscillation period equal to half of the Fermi wavelength $\lambda_{F}/2=30$ nm.

The results of Fig. 2(b) -- the $G$ map for both slits open -- were calculated as a sum of contributions of 16 incident subbands
given by separate scattering solutions (see the Theory section).
The contribution
of the incident subband $k$ to the SGM image when both the slits are open is proportional
to the probability density for the superposition of waves passing through one of the slits,
\begin{equation}
I_{l}\left(\boldsymbol{r}\right)=T_{l}\left|\psi_{\mathrm{qpc1}}(\boldsymbol{r})\Phi_{bs1}(\boldsymbol{r})\pm\psi_{\mathrm{qpc2}}(\boldsymbol{r})\Phi_{bs2}(\boldsymbol{r})\right|^{2},
\end{equation}
where the sign $\pm$ depends on the parity of the incident mode $l$. The parity is $(-1)^{l+1}$, where $l$ numbers
the energy subbands.
For only two input channel subbands at the Fermi level the SGM image is given by
\begin{eqnarray*}
I_{SGM} & = & I_{1}+I_{2}=\left(T_{1}+T_{2}\right)\left(\left|\psi_{\mathrm{qpc1}}\phi_{bs1}\right|^{2}+\left|\psi_{\mathrm{qpc2}}\phi_{bs2}\right|^{2}\right)\\
 & + & (T_{1}-T_{2})\cdot2\Re\left\{ \psi_{\mathrm{qpc1}}\phi_{bs1}\psi_{\mathrm{qpc2}}^{*}\phi_{bs2}^{*}\right\},
\end{eqnarray*}
which can be written as
\begin{equation}
I_{SGM}=(T_{1}+T_{2})I_{\mathrm{no\, int}}+(T_{1}-T_{2})I_{\mathrm{int}},
\end{equation}
where
$I_{\mathrm{no\, int}}$ is the sum of single-slit images as given by Eq. (10), and the $I_{int}$ is responsible for the interference effects.
From the above formula it is clear that the interference signal vanishes  when $T_1\simeq T_2$.
Figures 3(b) and 3(c) show the probability densities for the lowest and second incident subbands,
calculated from wave function given by a sum and a difference of integrals of type (5), respectively.
For the symmetry reasons in the lowest subband we have a central
line of positive interference [Fig. 3(b)] for the lowest subband and a zero of the wave function for the second subband  [Fig. 3(c)].
In Fig. 3(e,f) we plotted the $G$ maps obtained for the two subbands as calculated from the solution of scattering problem.
The maxima of the probability density coincide with the extrema of $G$ maps.
The latter contains the interference fringes of $\lambda_F/2$ period which correspond to standing waves between the tip and the QPC
which were described above for a single QPC [cf. Fig. 2(a) and Fig. 3(d)].

Generalizing the description, for $M$ subbands in the input channel the SGM image is given as
\begin{equation}
I_{SGM}=I_{\mathrm{no\, int}}T+I_{\mathrm{int}}\Delta_{T},
\end{equation}
with
\begin{equation}
T=\sum_{i=1}^{M}T_{i}
\end{equation} and
\begin{equation}
\Delta_{T}=T_{1}-T_{2}+T_{3}-T_{4}+...+T_{M-1}-T_{M}.
\end{equation}
For the SGM image normalized to the summed transfer probability one obtains
\begin{equation}
\frac{I_{SGM}}{T}=I_{\mathrm{no\, int}}+I_{\mathrm{int}}\frac{\Delta_{T}}{T},
\end{equation}
with the interference term vanishing for large $M$ since $\Delta_{T}<<{T}$.
This explains the result of Fig. 2, with the pattern of the $G$ map for the two QPCs [Fig. 2(b)] as a simple sum of maps for
separate QPCs [Fig. 2(c)].

\begin{figure}[htbp]
\includegraphics[width=0.4\paperwidth]{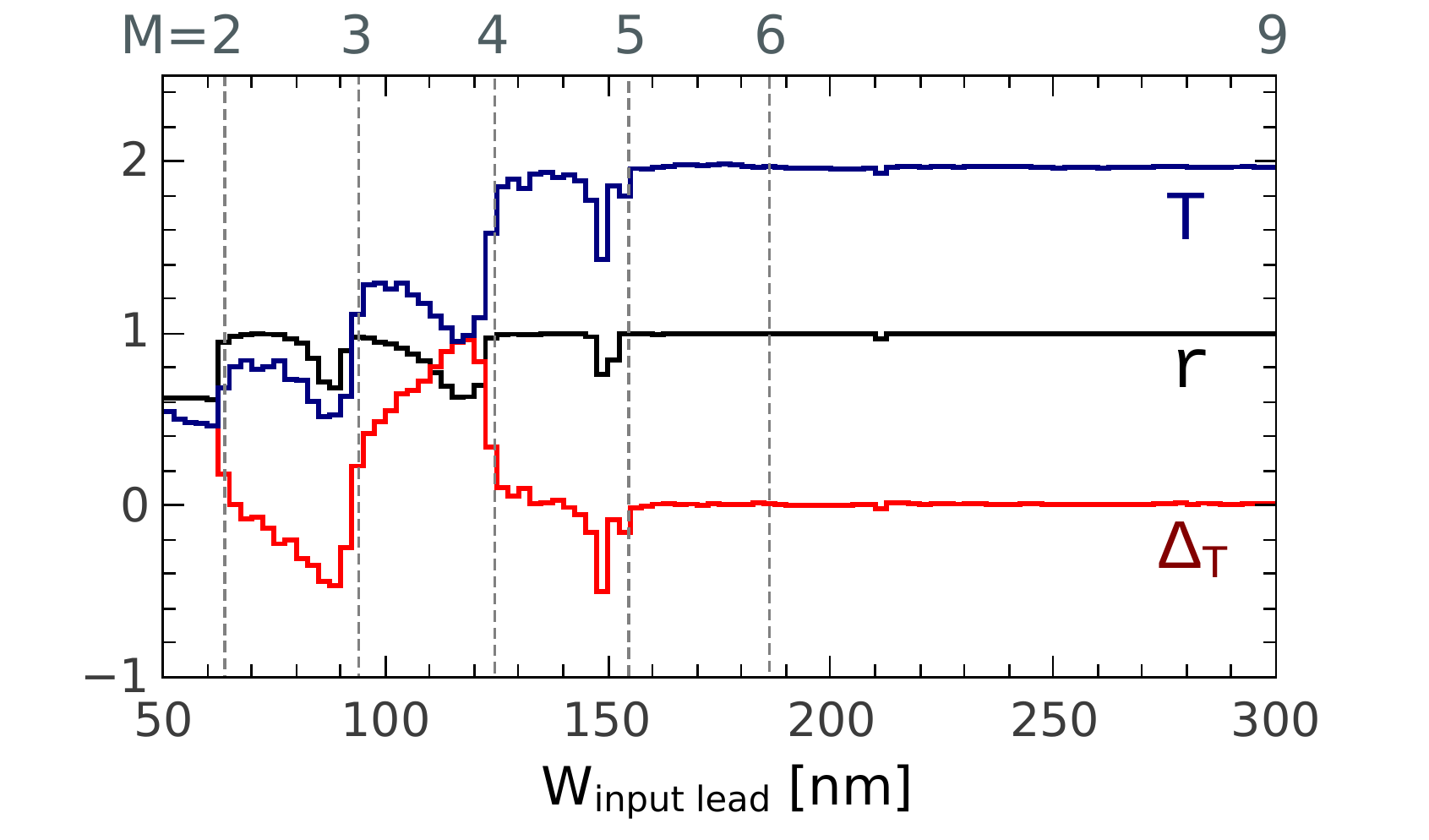}
\caption{\label{szkanszak} The number of incident subbands $M$ (top of the figure), the summed transfer probability $T$ [Eq. (15), blue line],
and the value of $\Delta T$ [Eq. (16)] as functions of the width of the input channel with fixed $E_F=6$ meV.
The black line ($r$) shows the correlation factor between the images of the probability density for both slit open,
and the sum of densities for separate slits. The value of $r=1$ indicates absence of the interference signal. The lower value of $r$
the larger the interference features.}
\end{figure}

The disappearance of the interference signal with $M$ is described in a quantitative manner in Fig. \ref{szkanszak}.
We keep  $E_F=6$ meV and vary the width of the input channel. The number of subbands $M$ is given
on top of the Figure. The summed transfer probability tends to $T=2$ (each slit opens a single channel) for large $M$.
The value of $\Delta_T$ vanishes for $M>4$.
For quantitative evaluation of the interference image we calculated the cross correlation coefficient $r$ \cite{kolas}
between i) the scattering probability density for both slits open and ii) the sum of densities for separate open slits
(the black line in Fig. \ref{szkanszak}). The perfect correlation $r=1$ corresponds to
a complete absence of the interference features in the probability density map. The lower value of $r$ the more distinct
are the interference features.
We can see that
for $M>4$, $r$ becomes equal to 1 in consistence with the value of $\Delta_T$ tending to 0.
Interference features are present for lower number of subbands, but only for $M=1$ the value of $r$ does
not depend on the width of the input channel.

\begin{figure}[htbp]
\includegraphics[width=0.4\paperwidth]{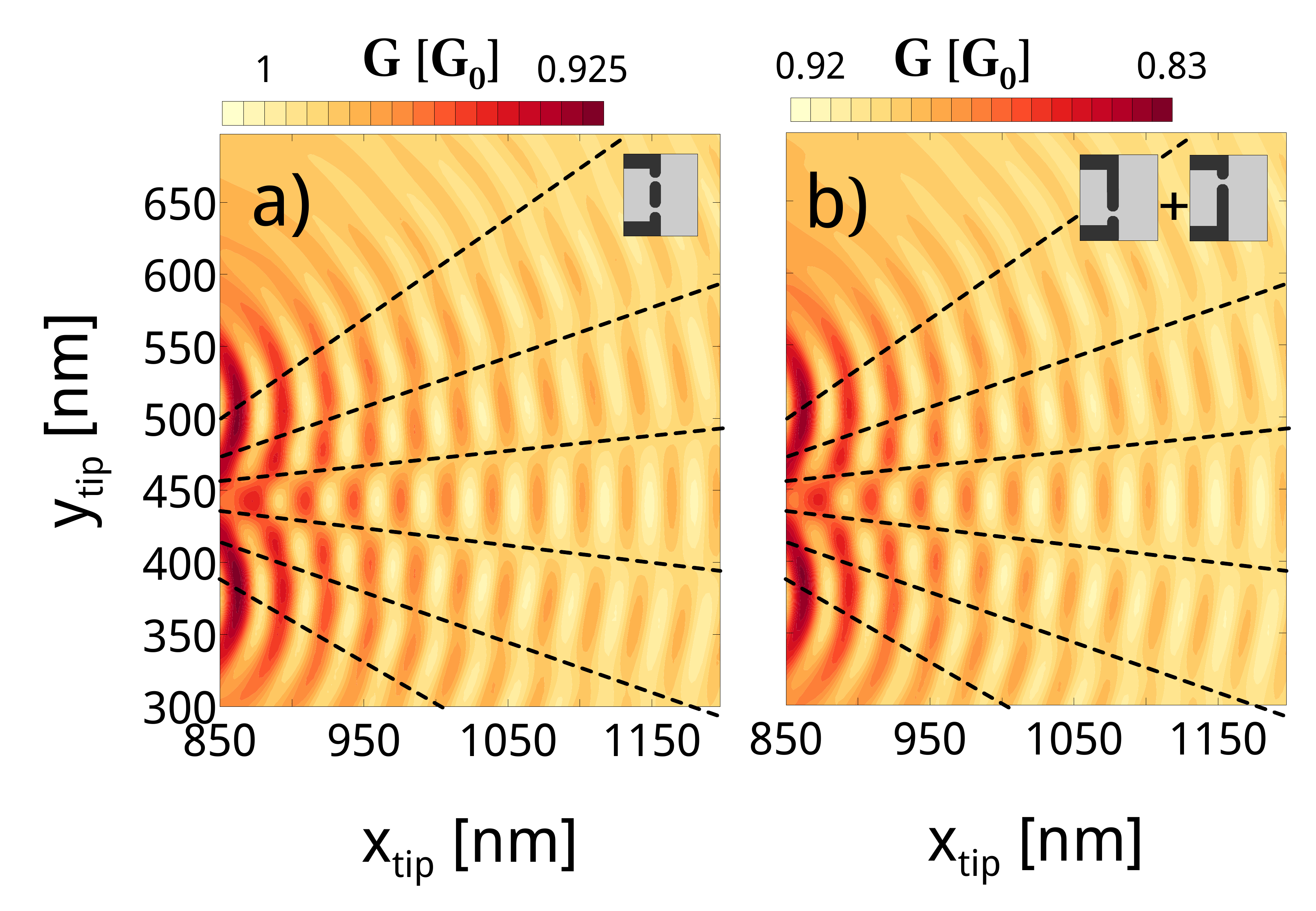}
\caption{\label{dodany} Same as Fig. 2(b) and (c) only for the QPCs shifted by 60 nm below the axis of the channel.}
\end{figure}

\begin{figure}[htbp]
\includegraphics[width=0.4\paperwidth]{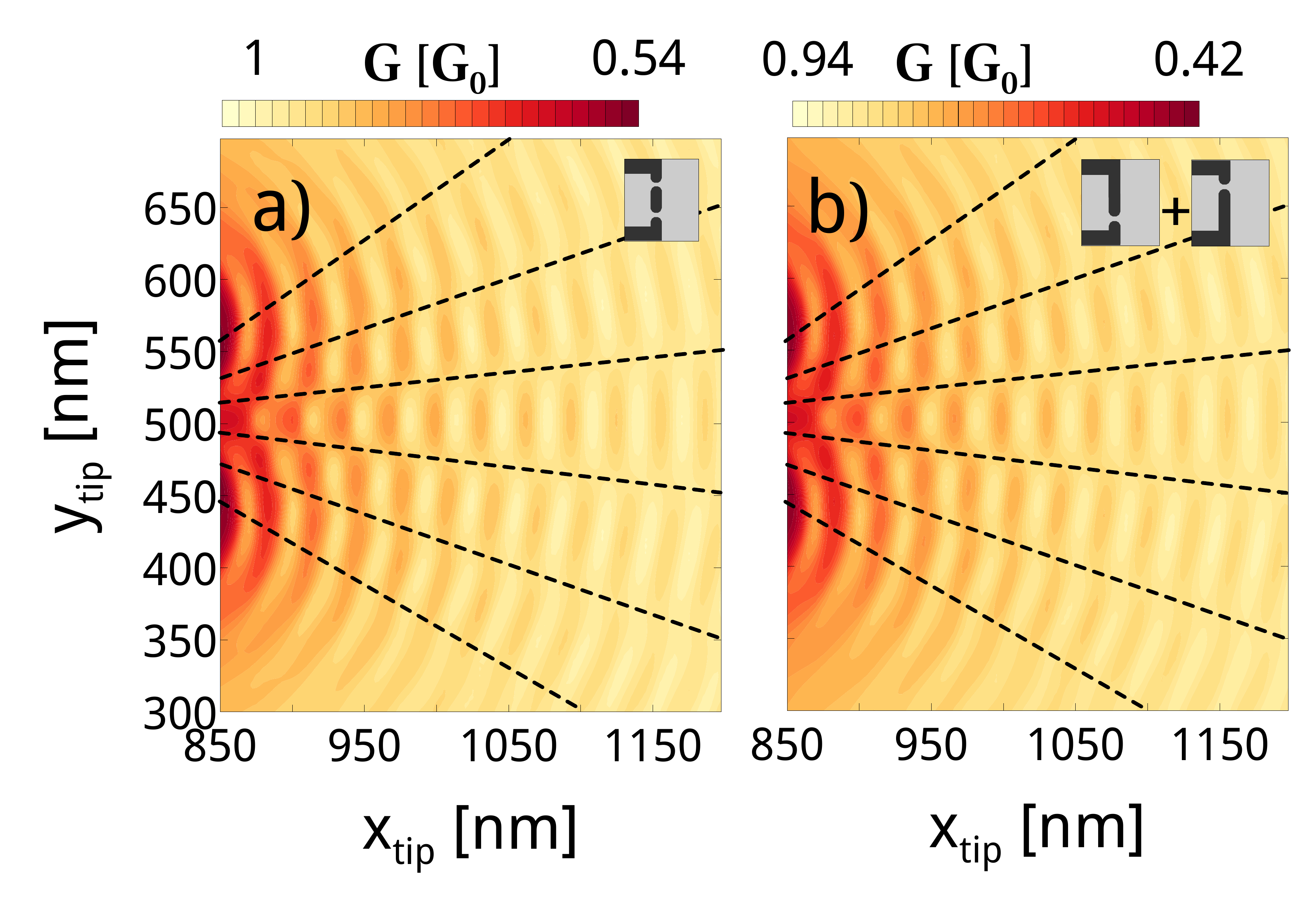}
\caption{\label{bigtip} Same as Fig. 2(b) and (c) only for $U_{tip}$ and $d_{tip}$ increased twice, to 10 mev and 20 nm, respectively}
\end{figure}

\begin{figure*}[htbp]
\includegraphics[width=0.8\paperwidth]{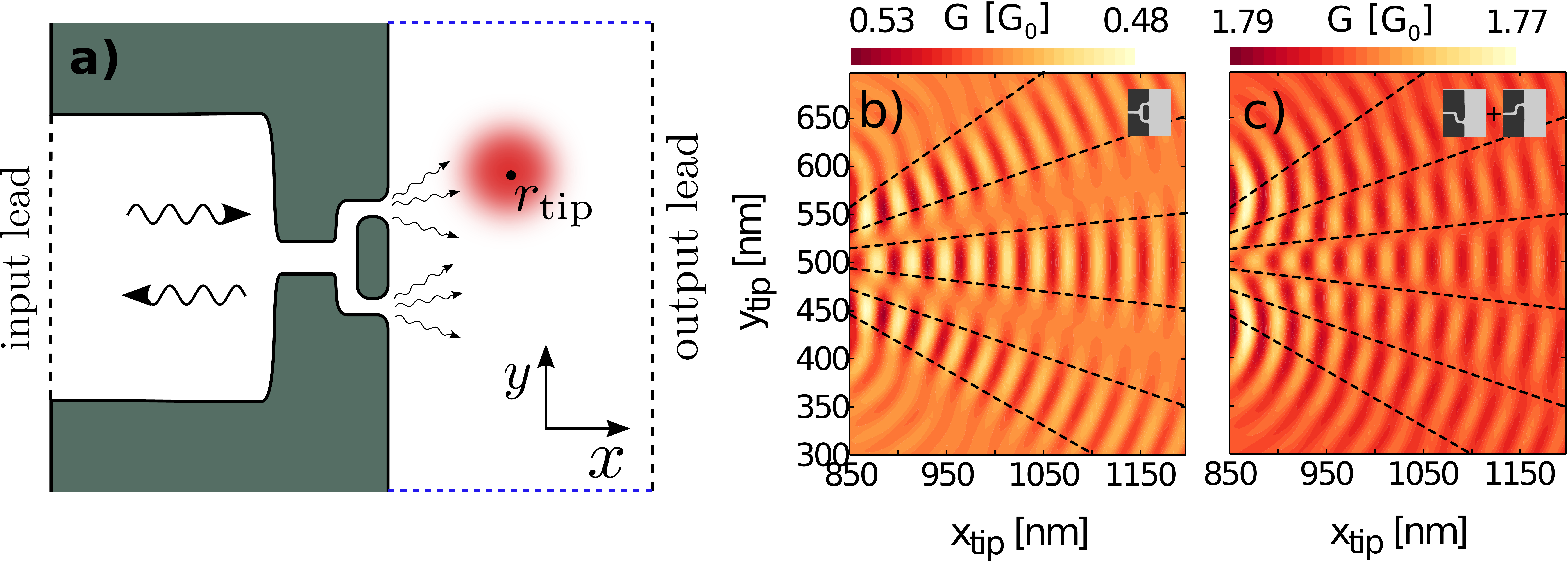}
\caption{\label{filtr} (a) Sketch of the system of double QPCs including
a channel acting as the symmetry filter. The width of channel is equal to the width of the opening $w=22$nm. (b) Conductance map for two QPC open. (c) Sum of conductance
maps of separate QPCs.}
\end{figure*}

Although the present interpretation is based on the symmetry of the incident eigenstates with respect to the axis of the input lead
we found that the absence of the Young interference is robust (see Fig. \ref{dodany}) against localization of the slits with respect to the axis.
The present discussion of the interference effects is independent of the parameters of the tip potential.
In Fig. \ref{bigtip} we present the conductance maps as obtained for the larger and wider tip potential with $d_{tip}=20$ nm, $U_{tip}=10$ meV, i.e.,
for both parameters increased twice as compared to the values used previously. The maps acquire a larger amplitude for the increased tip potential [cf. Fig.(b,c)], but $G$ for both slits
open still has the pattern as given by a simple sum of maps for separate slits.

The above discussion indicates that the absence of
the Young interference for many incident subbands results from cancelation of terms for various contributions to the total conductance in terms of the Landauer approach.
The present finding of an absence of the interference in the case of several subbands is related to the
the suppression of  the Aharonov-Bohm conductance oscillations with the period of the flux quantum \cite{but}.
Reference \cite{but} indicated a { reduction} of the amplitude of the Aharonov-Bohm oscillation
which { decreases} with the number of subbands as $1/M$. On the other hand, in the present work
we demonstrated a complete {\it removal} of the interference effects already at $M=5$.

A way to preserve a clear interference signal in the $G$ map is to filter out all the contributions but one.
For that purpose we considered a channel of width $w=22$ nm that feeds both the QPCs  -- see Fig. \ref{filtr}(a).
The conductance maps calculated for both QPCs open and the 16 subbands incident in the input lead is given in Fig. \ref{filtr}(b).
The horizontal channel has subbands on its own and at the Fermi energy its transmits current only in the lowest one. The total conductance
map of Fig. \ref{filtr}(b) with the radial features and maxima collimated into three beams is similar to the one obtained above
for the lowest subband contribution of Fig. 2(e). For comparison Fig. \ref{filtr}(c) shows the simple sum of $G$ maps obtained for a single QPCs open
with the checkerboard pattern encountered above in Fig. 2(b) before the filter channel was introduced.

\section{Summary and Conclusions} We have discussed the SGM imaging of the Fermi level wave function interference with the conductance maps obtained
by the scanning gate microscopy.
We solved the coherent scattering problem and demonstrated that although the double slit interference is present
for each of the incident subbands separately the signatures of the interference disappear when the total conductance
is evaluated by the Landauer formula. We explained the absence of the double slit interference
using symmetry arguments for the incident wave functions. A geometry in which the Young experiment can
be performed by SGM was indicated.




{\bf Acknowledgments}
This work was supported by National Science Centre
according to decision DEC-2012/05/B/ST3/03290, and
by PL-Grid Infrastructure.


\begin{thebibliography}{1}
\bibitem{actual} The actual computational box at right of the QPCs is 1$\mu$m by 1$\mu$m.
\end{thebibliography}
\end{document}